\documentclass[twoside,agupp]{aguplus} 

\usepackage{graphicx}
\lefthead{SOFIA AND LI}

\righthead{SOLAR VARIABILITY AND CLIMATE}

\received{October~02,~2000}
\revised{}
\accepted{}

\paperid{}

\cpright{AGU}{1998}

\ccc{0148-0227/97/97GL-00000\$05.00}

\authoraddr{S. Sofia and L.H. Li,
Department of Astronomy,
Yale University, 
New Haven, CT 06520-8101.  (e-mail: 
sofia@astro.yale.edu;
li@astro.yale.edu)}

\begin{document}

\title{Solar variability and climate}

\author{Sabatino Sofia and Linghuai H. Li \altaffilmark{1}} 

\affil{Department of Astronomy,
Yale University, 
New Haven, CT 06520-8101}

\altaffiltext{1}{Also at Purple Mountain Observatory,
 Chinese Academy of Science, Nanjing, Jiangsu 210008, China.}

\begin{abstract}
Recent precise observations of solar global parameters are used to calibrate an upgraded solar model which takes into account magnetic fields in the solar interior. Historical data about sunspot numbers (from 1500 to the present) and solar radius changes (between 1715 and 1979) are used to compute solar variability on years to centuries timescales. The results show that although the 11 year variability of the total irradiance is of the order of $0.1\%$, additional, longer lived changes of the order of $0.1\%$ may have occurred in the past centuries. These could, for example, account for the occurrence of climate excursions such as little ice ages.
\end{abstract}

\begin{article}

\section{Introduction}

Because of the steady increase of the concentration of greenhouse gases in the Earth's atmosphere due to anthropogenic activity, global warming is potentially the most serious environmental threat of our time. In order to assess the magnitude of the threat, we must understand the sensitivity of the climate system to a variety of external driving processes. In particular, the magnitude of the solar variability component of climate change, potentially very significant, is still relatively uncertain. 

One way of assessing the effects of a given driver of climate is by trying to fit the observations (e.g. reproducing the air temperature variations of the last 150 years \citep{JWW86,HL88} or sea surface temperature \citep{R91}) with, and without the driver in question operating. For the case of the Sun, we can assume that it does not change, and adjust all the other parameters to reproduce the temperature record. Subsequently, a given model of the solar variability is assumed, and then all the parameters are re-adjusted to reproduce the same temperature record. Obviously, if the solar variability contributes to the observed trend, the model sensivisity to the other processes that were assumed to produce the trend on their own must decrease. Similarly, if the solar component varies in antiphase with the observed trend, the sensitivity to the other processes must be increased. Andronova and Schlesinger carried out such a calculation \citep{AS00}. They used two models for the solar variability.  One, constructed by \citet{LBB95}, henceforth denoted by L, was based on sunspot areas and (when available) locations, He 1083 nm emission, group sunspot numbers, and Ca emission from the Sun and solar analogues.  The other model, developed by \citet{HS93}, (henceforth called HS), was based on the fraction of sunspot area occupied by the penumbra, equatorial rotation rate, length and decay rate of the solar cycle, and the mean level of solar activity.  Upon including solar variations, the model sensitivity to greenhouse gases changed drastically, decreasing by 53 percent for the HS solar variations, and by 44 percent for the L solar variation. Obviously, if either model provides an approximately accurate representation of solar variability over the last 150 years, the solar input to climate variability is very important.

\section{Our Solar Variability Model}

The problem with the models of solar variability referred to above is that they are based purely on empirical correlations which are not supported independently either by theory, or by direct, quantitative observations. As a consequence, we do not know how credible these extrapolations are.  In fact, although the explanation that the short term (days to months) variability of the total irradiance is largely due to active regions (spots and faculae) is beyond question, the precise origin of the 11 year modulation is still uncertain. The most popular explanation is that the modulation is similarly caused by contributions from active regions plus the magnetic network, which are known to vary with the activity cycle. In this explanation, the photospheric background, and everything below the photosphere, remain unchanged.  Alternatively, Li and Sofia \citet{LS00} propose a different explanation in which the 11 year modulation of the irradiance is due to structural adjustments of the solar interior produced by variations of an internal magnetic field located at about $0.96 R_{\sun}$. These adjustments modify the energy flow from the interior to the surface, and affect the entire luminosity of the Sun. This would imply that both the structure of the Sun and the photospheric temperature should vary over the activity cycle, and various observations of the solar photospheric temperature \citep{GL97a,GL97b}, solar oscillation frequencies \citep{LW90} and solar radius \citep{EKBS00,ACT00} support this explanation.

A variable magnetic field in the solar interior adds a variable magnetic component to pressure and internal energy not considered in standard models. In addition, the field modifies the transfer of convective energy. As a consequence, the structure of the solar interior changes (albeit slightly) from the field-free state, and so do all global parameters, such as radius and luminosity (hence, effective temperature). The formulation of this model upgrade is developed in Lydon and Sofia \citet{LS95}, and updated in Li and Sofia \citet{LS00}. Model calculations show that all conventional (nonmagnetic) global parameters are sensitive to the location, intensity, distribution and orientation of the solar interior magnetic field, and this enables us to use the observed cyclic variations of solar radius, luminosity and effective temperature to determine cyclic variations of those magnetic properties.

\begin{figure}[t]
\figurewidth{8cm}
  \includegraphics[width=8cm]{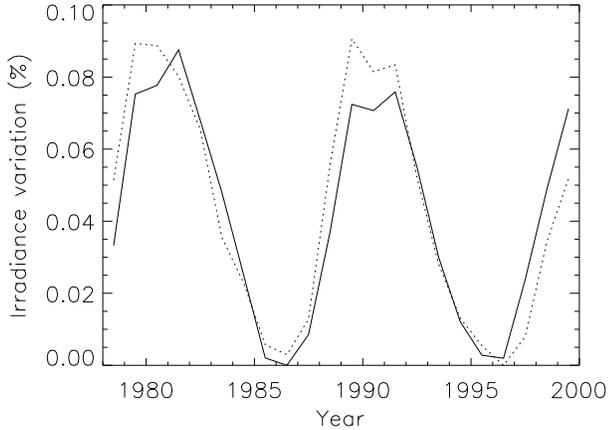}
 \caption{Comparison between the measured (solid curve) and calculated (dashed curve) solar irradiance variations.
}
\label{fig:ir}
\end{figure}

It is well known that the even splitting coefficients vary with solar activity cycle \citep{LW90,BJT99,HKH99}. Since magnetic fields contribute to the even splitting coefficients, and the field causes the activity as it buoys to the surface, we can relate the maximum magnetic field $B_m$ to the sunspot number $R_Z$ in the following way:
\begin{equation}
  B_m=B_0 \{ A+[1+\log(1+R_Z)]^5 \}. \label{eq:bm}
\end{equation}
The magnetic field inferred  from the solar oscillation frequencies observed around 1996 was equal to about 20 kG, and it peaked at $r\approx 0.96 R_{\sun}$ \citep{ACT00}. To match the even splitting coefficients, the magnetic field distribution in the solar interior is assumed to have a gaussian-like profile, and to be purely toroidal. Since the location of the field is fixed at around $r=0.96 R_{\sun}$ by more accurate helioseimology, the more uncertain observation of the solar radius variation is not needed. Consequently, $B_0$, $A$, and the profile width can be determined by fit for the simultaneously-observed cyclic variations of solar irradiance \citep{WH91,FL98} and effective temperature \citep{GL97a,GL97b}.

\begin{figure}[t]
\figurewidth{8cm}
  \includegraphics[width=8.cm]{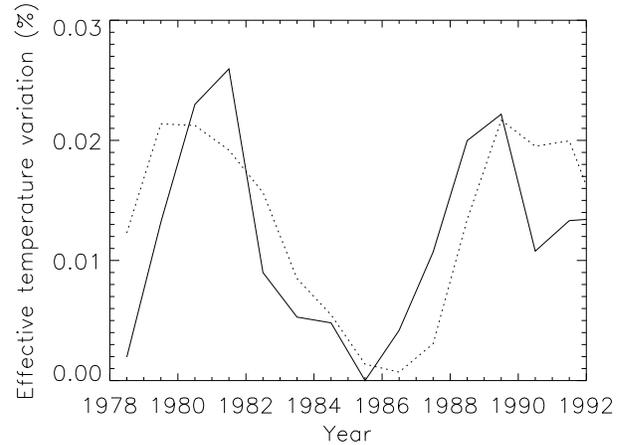}
\caption{Comparison between the measured (solid curve) and calculated (dashed curves) solar photospheric temperature variations.
}
\label{fig:te}
\end{figure}

Since 1978, several space-borne radiometers have monitored solar irradiance and provided us with daily irradiance data of unprecedented precision \citep{FL98}. In our model calculations we use one year as the timestep, so we use yearly averages of the total solar irradiance. The standard variances range from 0.1 to 0.5 Wm$^{-2}$. The measured irradiance varies almost in phase with the solar activity cycle with a relative amplitude of about $0.1\%$ (solid curve in Fig.~\ref{fig:ir}). Gray and Livingston [1997a,b] measured solar effective temperature variations over the 1978-1992 interval by monitoring the neutral carbon $\lambda5380$ line in the solar flux spectrum. They found that the temperature varies about $1.5\pm0.2$ K during a cycle (solid curve in Fig.~\ref{fig:te}, also nearly in phase with the cycle. Since this line has a high excitation potential (7.68 eV), it forms deep in the photosphere. As a result, the effective temperature measured in this way is actually a photospheric temperature, thus it is free of the effects of changing active regions and network. The best fit for the observed cyclic variations of solar irradiance and temperature is obtained by setting $B_0=90$ G, $A=190$, and $0.06 R_{\sun}$ for the magnetic field profile width. Our model fits the irradiance observation well ($\chi^2_1=4$), and the temperature a little less well ($\chi_2^2=43$). To obtain the best fit we minimize $\chi^2=\chi_1^2+\chi_2^2$. This best fit model is also consistent with helioseismology \citep{ACT00}. For example, using Eq.~(1) with $B_0=90$ G, $A=190$ and $R_Z=8.6$ in 1996, we find $B_m \approx 20$ kG. Of course, it peaks at $r=0.96R_{\sun}$ as required by helioseismology \citep{ACT00}. The resulting W parameter is almost constant and equal to about $2\times 10^{-5}$.

\begin{figure}[t]
\figurewidth{8cm}
  \includegraphics[width=8.cm]{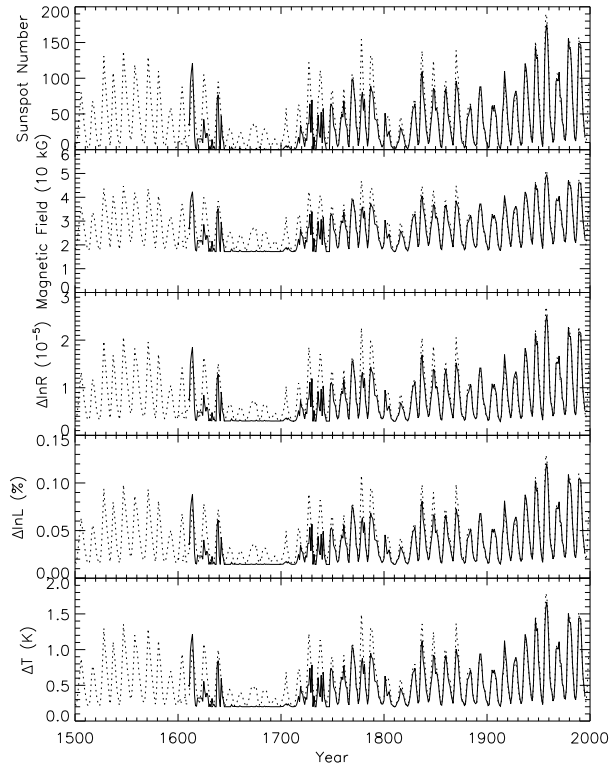}
\caption{Solar variability in the past five centuries (dotted curve corresponds to the Z\"{u}rich sunspot number $R_Z$) and in the past four centuries (solid curve conrresponds to the group sunspot number $R_G$).
kG stands for kilo-Gauss, $L$ for total solar luminosity, $T$ for solar effective temperature, $R$ for solar radius.
}
\label{fig:yearly}
\end{figure}

\section{Extrapolation of the Activity Cycle Timescale Variability}

After calibrating our model by using the recent precise data for solar irradiance during the period from 1978 to 1999, and for effective temperature during the period from 1978 to 1992, we can extrapolate the solar irradiance back during the period when the annual sunspot numbers are available.

Surely the best-known features of the Sun are the sunspots and the regular cycle of the solar activity, which waxes and wanes with a period of about 11 years. Naked-eye observations of sunspots have been made in China since before A.D. 300, and telescopic observations in Europe have been made since A.D. 1610. The discovery of the sunspot cycle by Schwabe in 1843 led Wolf in 1847 to 1851 to the concept of universal sunspot numbers, which in 1868 he extended backward annually to 1700. At Z\"{u}rich, Wolf calculated magnitudes of each cycle from 1749, and the central dates or epochs of maxima and minima back to about 1610. Early aurorae or Northern lights provide clues to the dating of early sunspot cycles. The printed catalogs and additional ``Spectrum of Time'' material \citep{S48,S67,E76,S61} have made it possible to extend Wolf's tables \citep{W51} back into earlier centuries.

On scrutinizing the historical sunspot data, Schove \citet{S83} obtained an annual sunspot number table starting from A.D. 1500, as shown in the top panel of Fig~\ref{fig:yearly} (dashed line). The gap from 1645 to 1715 corresponds to the so-called Maunder minimum. The second panel from the top in Fig.~\ref{fig:yearly} shows the maximum magnetic field strength calculated by Eq.~(\ref{eq:bm}) with the best fit parameters. Obviously, the magnetic field strength equals to about 17 kG even if the sunspot number is zero (e.g. in 1810), since helioseismology requires 20 kG for 1996. The last three panels show the cyclic variations of solar luminosity, radius and effective temperature, respectively. We also use the more precise group sunspot numbers \citep{HS98} to recalculate solar variability in the past, as shown by the solid curves in Fig.~\ref{fig:yearly}. Obviously, both results have the same trend. Because luminosity changes result in solar irradiance changes and the relationship is linear, the (relative) solar irradiance change is the same as the (relative) luminosity change. Detailed tabulations of these data will be provided on request.

\section{Additional Variability Mechanisms}

As we can see from Fig.~\ref{fig:yearly}, the maximum variability of the solar radius is about $2\times 10^{-5}$, or 0.02 arc~s.  Although this variation is in agreement with the most recent determination of the cycle radius variations obtained from the MDI experiment on SOHO \citep{EKBS00}, it is much smaller than the radius changes determined from historical data over the last 2 centuries. In our view, the most reliable historical data sets from which solar radius changes can be determined are the duration of total eclipses measured near the edges of totality. From them, changes of the order of 0.5 arc~s have been detected.  In particular, a change of $0.34$ arc~s between 1715 and 1979 (Dunham et al 1980), a change of $0.5$ arc s between 1925 and 1979 \citep{DSFHM80}, and no change between 1979 and 1976 \citep{SDDF83}, were detected. If such changes are real, what could cause them? What are the corresponding solar irradiance changes?

\begin{figure}[t]
\figurewidth{8cm}
  \includegraphics[width=8.cm]{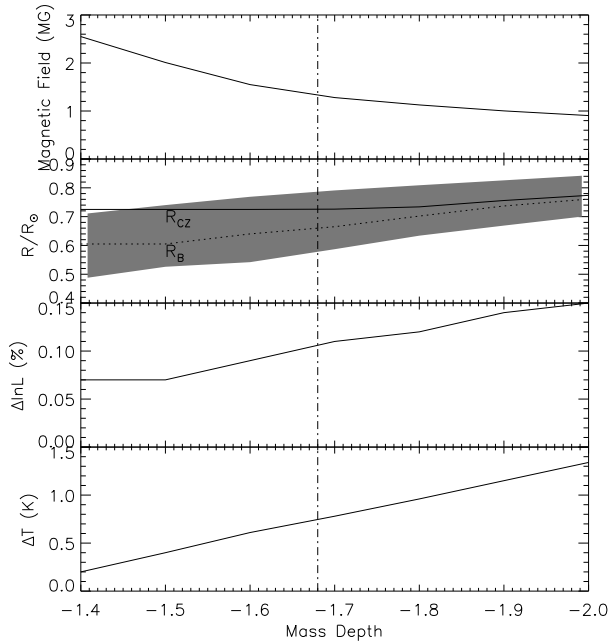}
\caption{Solar variability corresponding to the probable change in the solar radius between 1715 to 1979.
$R_{\mbox{\scriptsize{CZ}}}$ is the location of the base of the convection zone, while $R_B$ is the location of maximal magnetic field. MG stands for Million Gauss. The mass depth is defined as $\log(1-M/M_{\sun})$ by the mass coordinate M. The smaller the mass depth, the closer to the surface of the model sun.
}
\label{fig:century}
\end{figure}

If we use the magnetic field location required to produce the 11 year cycle variability, we find that it is impossible to produce a 0.5 arc~s radius variation even if we apply an unreasonablely strong magnetic field. However, our model shows that the deeper the location of a magnetic field and the more intense the magnetic field, the larger the resulting radius change. The physical cause is obvious: the stronger the magnetic field, the larger the contribution of magnetic pressure to the fixed total pressure required by hydrostatic equilibrium. Consequently, the gas pressure decreases accordingly and the gas will expand to increase the solar radius. However, gas expansion is limited by the gas density. Therefore, we compute the magnetic field required to produce the detected radius change between 1715 and 1979 as a function of mass depth, as displayed in the top panel in Fig.~\ref{fig:century}. It is well known that a strong magnetic field will cause a change of location of the boundary between the convective and the radiative region \citep{LS95}. The second panel from the top in this figure shows how the convection boundary $R_{\mbox{\scriptsize{CZ}}}$ varies with the applied magnetic field (solid curve), and how the location of the maximal magnetic field, $R_B$, varies with the mass depth (dashed curve). The shadowed region indicates the half-width of the required magnetic field. Of particular relevance are the values corresponding to the base of the convection zone, as indicated by the dot-dashed line in this figure, since all conventional dynamo models locate the process precisely at that depth. There, the magnetic field required to cause a 0."34 change of the solar radius is 1.3 million G, and the resulting luminosity variation is 0.12 percent (the third panel of Fig.~\ref{fig:century}), which is almost equally due to the variation of effective temperature (the bottom panel) and radius, since the radius variation contributes $2\times \Delta\ln R=0.07\%$. These values are interesting for producing significant climate change if the solar variations are sufficiently long lasting, and for not grossly contradicting what we know about the Sun, excepting a value for the magnetic field that is larger than we are comfortable with.

\section{Results and Discussion}

According to the picture presented in this paper, the solar variations of years to centuries are due to variations of magnetic fields at two different depths in the solar interior.  The 11 year cycle would arise from magnetic fields of the order of 20-40 kG, located at or above $0.96 R_{\sun}$.  At the same time, there is a stronger field located at just below the base of the convection zone that undergoes changes on longer timescales. In fact, this field affects where the base of the convection zone is located. The temporal characteristics of the deeper field is still unknown, since the radius data currently available have a scant number of data points, poorly distributed in time.  The amplitude of the variations of the solar luminosity produced by the shallower field is of the order of $0.1\%$, while the deeper field can produce changes of the same order of magnitude. The solar luminosity when both magnetic components are strong would be about $0.2\%$ higher than when both are weak.

For instance, the luminosity deficit caused by the shallow field at the time of the Maunder minimum is about $0.1\%$. To cause the decrease of the radius observed between 1715 and 1979, the magnetic field at the base of the convection zone had to decrease from about 1.3 MG to about 600 kG. Since 1979 is at the maximum of Cycle 21, the corresponding magnetic field at the minimum is smaller than one half of the field at the maximum. We thus infer that the magnetic field at the minimum is smaller than 300 kG, which is in agreement with helioseismology \citep{ACT00}. The luminosity due to the deep field in 1715 is higher than that in 1979 by about $0.1\%$, but is almost the same as that in 1925. The Maunder minimum terminated around 1715, when both shallow and deep magnetic field components are, according to our model results, strong. We know that the shallow magnetic field was weak during the Maunder minimum, because of the small sunspot numbers during that period. If we assume that the deeper component was also weak, the solar luminosity during the Maunder minimum may have been about $0.2\%$ lower than that after 1715. Such a large irradiance change may account for the transition between the little ice age and the following warm period.

The solar radius change found to occur between 1925 and 1976 (or 1979) brings up an additional issue. The implication is that the solar luminosity in 1925 was about $0.1\%$ higher than that in 1976 (or 1979). Of course, the climate consequences of this higher luminosity period depend strongly on the duration of this phenomenon, which at present is unknown.  However, there is a possibility that the solar effect alone would have caused some cooling, and that the global warming caused by the increase of the concentration of greenhouse gases could be more significant than what has been estimated ignoring the solar effect. Because of the complexity of the climate response to solar irradiance, and in view of the serious uncertainties in the current solar data base, we cannot, at present, draw any sweeping conclusions.  What we can conclude with some certainty is that solar variability could be a significant driver of climate change, and that the efforts made to date in addressing this issue are inadequate. The physical model that we have developed, with all its inadequacies (for example, it is unidimesional, whereas the problem is clearly three-dimensional) can constitute a framework which could form the basis for interpreting a variety of historical solar observations. 

\acknowledgments

This work was supported in part by NASA and NSFC (project 19675064).

{}
\end{article}

\end{document}